# *Spectral region optimization for Raman-based optical diagnosis of inflammatory lesions*

Luis Felipe das Chagas e Silva de Carvalho[1,2], Renata Andrade Bitar[3], Emília Ângela Loschiavo Arisawa[3], Adriana Aigotti Haberbeck Brandão[1], Kathia Maria Honório[4], Luiz Antônio Guimarães Cabral[1], Airton Abrahão Martin[3], Herculano da Silva Martinho[2*], Janete Dias Almeida[1]

[1]Departamento de Biociências e Diagnóstico Bucal,

Faculdade de Odontologia de São José dos Campos – FOSJC, Brazil

[2]Centro de Ciências Naturais e Humanas, Universidade Federal do ABC – UFABC,

09090-400, Santo André, São Paulo, Brazil

[3]Laboratório de Espectroscopia Vibracional Biomédica,

Instituto de Pesquisa e Desenvolvimento - IP&D,

Universidade do Vale do Paraíba - UniVap, Av. Shishima Hifumi,

São José dos Campos, São Paulo, Brazil

[4]Escola de Artes, Ciências e Humanidades, Universidade de São Paulo, Av. Arlindo

Bettio, 1000, 03828-000 São Paulo - SP, Brasil

E-mail: luisfelipecarvalho@hotmail.com, rabc@univap.br, mirela@univap.br, adriana@unesp.br, kmhonorio@usp.br, cabral@fosjc.unesp.br, amartin@univap.br, herculano.martinho@ufabc.edu.br, janete@fosjc.unesp.br

## Abstract

FT-Raman Spectroscopy was applied to identify biochemical alterations existing between inflammatory fibrous hyperplasia (IFH) and normal tissues of buccal mucosa. One important implication of this study is related to the cancer lesion border. In fact, the cancerous – normal border line is characterized by the presence of inflammation and its correct discrimination would increase the accuracy in delimiting the lesion frontier. Seventy spectra of IFH from 14 patients were compared to 30 spectra of normal tissue from 6 patients. The statistical analysis was performed with Principal Components Analysis and Soft Independent Modeling Class Analogy methodologies. After studying several spectral ranges it was concluded that the best discrimination capability (sensibility of 95% and specificity of 100%) was found using the 530 – 580 $cm^{-1}$ wavnumbers. The bands in this region are related to vibrational modes of Collagen aminoacids Cistine, Cysteine, and Proline and their relevant contribution to the classification probably relies on the extracellular matrix degeneration process occurring in the inflammatory tissues.

**1. Introduction**

The Raman-based optical biopsy methods have been presented as very promising non-invasive tools for precocious diagnosis of several pathologies. Actually, technological limitations need be overwhelmed in order to improve the sensibility and specificity of these methods for wide clinical applications. One of these limitations concerns the influence of inflammatory infiltrates on the correct pathology discrimination. Some recent works had pointed out that important early pathological states could be misdiagnosed due to the presence of inflammation[1-3].

Cancer affects thousands of people worldwide and is responsible for many deaths. A large rate of optical-biopsy studies on breast, skin, lung, and brain cancers had been performed[4-6]. However, there are few studies concerning oral cancers in literature[7,8]. An animal experimental model performed with 21 hamsters by Oliveira *et al*[7] showed that FT-Raman spectroscopy could differentiates normal, dysplastic, and oral squamous cell carcinoma tissues. Malini *et al*[3] applied Raman spectroscopy to differentiated normal, premalignant, inflammatory and cancer oral tissues. The authors were just able to discriminate normal and altered tissues. Venkatakrishna *et al*[8], studying 37 samples of cancer and 12 of normal oral tissues, concluded that Raman spectroscopy can differentiate them with 85% sensibility and 90% specificity.

Inflammatory fibrous hyperplasia (IFH) is a non-neoplastic benign lesion of the oral mucosa.[1] Its origin relies on some kind of low intensity chronic trauma as wearing an ill-fitting full or partial or fractured denture prosthesis, rubbing on fractured teeth with sharp edges, diastemas, improper oral hygiene[9]. The oral epithelium shows epithelial changes induced by the inflammation of the underlying lamina propria. In some situations, the epithelial changes are similar to epithelial dysplasia seen in a premalignant lesion[10]. IFH is more prevalent in women in the fifth decade of life[9]. The histological confirmation of the clinical diagnosis involves quite subjective inherent factors which limits the sensitivity for detection[9,10].

In order to characterize non-cancerous inflammatory oral tissues it was performed in the present work a comparative study between normal and IFH oral tissues. The main objective is to find the minimum spectral range enabling correct diagnosis with help of Principal Components Analysis (PCA) and Soft Independent Modeling of Class Analogy (SIMCA). One important implication of this study is related to the cancer lesion border. In fact, the cancerous – normal border line is characterized by the presence of inflammation and its correct discrimination would increase the accuracy in delimiting the lesion frontier.

**2. Material and Methods**

This research was carried out according to the ethical principles established by the Brazilian Healthy Ministry and was approved by the local ethical research committee 067/2006/PH-CEP. Patients were informed concerning the subject of the research and gave their permission for the collection of tissue samples.

**2.1. Sample preparation**

Samples of 14 patients diagnosed as IFH and 6 of normal tissues (NM) were obtained from biopsies performed at the Department of Bioscience and Oral Diagnosis – UNESP/BRAZIL. The tissue samples were identified and immediately snap frozen and stored in liquid nitrogen (77K) in cryogenic vials prior to FT-Raman spectra recording.

**2.2. FT-Raman Spectroscopy**

The Raman spectra were measured on 5 different points (A1-A5) as indicated on Fig.1. These resulted on 70 spectra of IFH and 30 spectra of NM. Soon after the procedure, all samples were fixed in 10 % formaldehyde solution for further histopathological analysis. A Bruker RFS 100/S FT-Raman spectrometer was used with a Nd:YAG laser operating at 1,064 nm as excitation light source. The laser power at the sample was kept at 230 mW and the spectrometer resolution was 4 cm$^{-1}$. Each spectrum was recorded with 300 scans. For FT-Raman data collection, all samples were brought to room temperature and kept moistened in 0.9 % physiological solution to preserve their structural characteristics, and placed in a windowless aluminum holder for the Raman spectra collection. We noticed that the chemical species present in the physiological solution ($Ca^{2+}$, $Na^+$, $K^+$, $Cl^-$, and water) do not have measurable Raman signals and their presence does not affect the spectral signal of the tissues.

**2.4. Histopathological analysis**

The histopathology of NM samples showed normal epithelium, lamina propria with appearance of normality, and collagen fibers arranged in wavy bundles with typical cellular components (Fig. 2a). IFH tissues (Fig. 2b) presented epithelial changes as hydropic degeneration, exocytosis, spongiosis, acanthosis and epithelial hyperplasia of cones.[Shafer] The collagen fibers presented thick and irregular shapes. The diffuse inflammatory infiltrate is predominantly mononuclear sometimes with blood vessels congestive. Depending on the relative amount of inflammatory cells the infiltrate could be classified as mild, moderate, or intense infiltrate.[Shafer]

**2.3. Data analysis**

In this work, some pattern recognition methods such as PCA and SIMCA were employed to analyze the data set and to obtain the relationship between Raman spectra and the two classes of collected samples. Before employing these methods, the spectral data were preprocessed (baseline corrected and normalized). Afterwards, all variables were mean-centered. The PCA and SIMCA analyses were carried out by using the Pirouette software (Infometrix Inc. (2002) Pirouette 3.11, Woodinville, WA). All analyses were investigated using cross-validated leave-one-out (LOO) method.

*Soft Independent Modelling of Class Analogy (SIMCA)*

In SIMCA method a PCA model is constructed for each sample class according to the position and distribution of the compounds in the row space.[Beebe, Sharaf] Consequently, a multidimensional (determined by the number of PCs necessary to describe the class) box is built for each class (this means that the shape and position of the samples in the classes are taken into account) and the limits of the boxes are defined according to a certain level of confidence. The classification of a test sample is achieved by determining which space the sample occupies and if it can be a member of one, more than one or none of the class (boxes). The number of principal component of each class is determined maximizing the sensibility and specificity [Beebe, Sharaf]. The main advantage of SIMCA over other classification methods is its ability to detect outlier samples [Beebe, Sharaf].

**3. Results and Discussion**

Figure 3 shows the Box Plot for the normal (Fig. 3a) and IFH (Fig. 3b) data. The black lines correspond to the average of the data while the vertical gray ones are the region between the first and third quartiles. The assignment of the main Raman bands is present on Table 1. The rectangular boxes on Fig. 3 indicate the spectral regions (close to 574, 1100, 1250-1350, and 1500 cm$^{-1}$) with biggest intra-group variations. According to Table 1, these bands are related to $CO_2$ rocking, CC stretching, Amide III / $CH_3$, $CH_2$ Twisting, $CH_2$ Bending and C = C stretching which are primarily related to proteins such as collagen. Actually, the IFH group presented lesser intra-group variation than the NM one. This fact could be related to the acanthosis process in IFH epithelium. In this process the thickness of the epithelium increases due to the growing of the spinous layer. This type of tissue has a more homogeneous composition than the connective one which implies greater similarity of the Raman spectra within the IFH group than the NM one. This reinforces the accuracy of the FT-Raman spectroscopy when validated by histopathological analysis.

Comparing the NM (Fig. 2a) and IFH (Fig. 2b) spectra one could state that the main differences were observed in the 1200 (C-C aromatic/DNA), 1350 ($CH_2$ Bending/Collagen 1) and 1730 cm$^{-1}$ (Collagen III) regions. These bands appeared less intense in the IFH group.

The change in the DNA band is associated to the increased proliferation of inflammatory cells (neutrophils, macrophages and lymphocytes) in inflammatory area [1]. This change, may also be related to the increase in the production of collagen fibers, due to the increased number of fibroblasts in inflammatory tissue or increasing the production capacity of collagen fibers that are intrinsic to each fibroblast (Wang et al, 2003).

The collagen bands intensity decreasing in the IFH group were closely related to the histopathological findings. The collagen was observed as parallel thin and delicate bundles of fibers in NM and thick and mature bundles of collagen fibers, arranged in different directions for IFH. The proliferation of inflammatory cells at the inflammation site such as lymphocytes, macrophages and neutrophils cause degradation of several macromolecules in the extracellular matrix, as shown by Séguier et al [Séguier, 2000] for gingivitis.

The technique of FT-Raman spectroscopy also identified the relationship between normal and inflammatory tissues through the PC's, demonstrating the effectiveness of the technique as a method of diagnosis assistant, or even complementary to histological examination [9].

As first step it was analyzed all spectral data covering 400 – 1820 cm$^{-1}$. As it was displayed in Figure 4 there was no significant discrimination among the two sample classes It is important to notice that 20 PCs sum 93.3% of information, where $PC_1$ = 59.5%, $PC_2$ = 18.2%, $PC_3$ = 4.8%, $PC_4$ = 2.7%, $PC_5$ = 1.7%, $PC_6$ = 1.4% and $PC_7$ = 0.71%.

After several attempts to achieve a better discrimination among the sample classes studied, the best separation was obtained with a small set of variables covering the region between 530 and 580 cm$^{-1}$, which corresponds to amino acids vibrations. The PCA results also show that the first six principal components ($PC_1$ to $PC_6$) describe 91.7% of the overall variance as follows: $PC_1$=79.47%, $PC_2$=2.98%, $PC_3$=2.65%, $PC_4$=2.65%, $PC_5$=2.06% and $PC_6$=1.90%. Since, almost all variance is explained by the first two PCs, their score plot is a reliable representation of the spatial distribution of the points for the data set. Figure 5 ($PC_2$ versus $PC_1$) presents the PCA score plot, indicating a good separation between NM and IFH samples.

The SIMCA analyses were performed using the three first PCs in the region between 530 and 580 $cm^{-1}$. The three-dimensional projection of the samples is shown on Figure 6 with the hyperboxes (small black points) representing two classes. The coordinates of the hyperboxes that determine the limits of the classes are obtained according to the standard deviations of the sample scores in the direction of each PC and states a confidence limit of 95% for the distribution of the classes.

Another way to analyze the SIMCA results is to observe the plot of the distances among sample classes, which are calculated according to the residuals of the samples when they are adjusted to the classes. In general, this plot is divided by two lines that represent the confidence limits (95%). The samples lying in the north–west quadrant (NW) belong to the y-axis class. Analogously, the samples in the south-east quadrant (SE) are members of the x-axis class only. Samples positioned in the south-west quadrant (SW) may belong to both classes, while the ones in the north-east quadrant (NE) belong to none. Figure 7 displays the plot of the distances among the classes of the samples studied in this work. From Figure 7 one can note that all normal samples are in the NW and SW quadrants, therefore they belong to Class 1. Otherwise, only five inflammatory spectra (labeled as HFI35A4, HFI53A2, HFI85A3, HFI77A3, HFI77A5 in the plot of Fig. 7) were classified incorrectly, i.e. these samples were classified into the quadrants occupied by the normal samples. However, only one inflammatory spectrum (HFI35A4) is located close to the normal class (the other four inflammatory samples near to the limit of classes). As these five spectra were taken near to the border lesion (points 2-5 of Fig. 1) probably they had normal tissue characteristics justifying the misclassification.

Thus, the sensibility and specificity obtained by the SINCA method were 95% and 100%, respectively. From these results one can consider this model suitable for a good discrimination among the sample classes.

One important point in this work concerns the very specific region (530 – 580 cm$^{-1}$) getting the better classification. As these bands are related to vibrational modes of Collagen aminoacids Cistine, Cysteine, and Proline their relevant contribution to the classification probably relies on the extracellular matrix degeneration process occurring the IFH. In this process, cytotoxic cells and proteolitic enzymes attach the fibroblasts and matrix macromolecules leading to a sudden and extensive breakdown of the collagen compound.[Seguier]

## 4. Conclusions

The analysis of the FT-Raman spectra of the NM an IFH buccal mucosa indicated that the SIMCA method presented a powerful discriminating capability (sensibility of 95% and specificity of 100%) when using the 530 – 580 cm$^{-1}$ spectral region. Thus, only using this spectral window is possible to discriminate normal and inflammatory tissues which is very useful information for accurate cancer border lesion determination. The existence of this very narrow spectral window enabling normal and inflammatory diagnosis had also useful implications for an "in vivo" dispersive Raman setup for clinical applications.

## Acknowledgments

The authors would like to thank the Brazilian agencies FAPESP and CNPQ for the financial support.

# TABLES

*Table 1 – Vibrational modes – Structural components (Movasaghi et al, 2007, Lyng et al, 2006)*

| BANDS (cm$^{-1}$) | VIBRATIONAL MODES | STRUCTURAL COMPONENTS |
|---|---|---|
| 530 - 580 | $CO_2$ rocking; S-S bridge | Cistine, Cysteine, Proline |
| 766 | C-C aromatic ring breathing | Pyrimidine |
| 850 | CCH deformation | Amino acids and polysaccharides |
| 959 | CC strething *α-helix* | Proteins |
| 1004 | C-C aromatic ring stretching | Phenylalanine (referred to collagen) |
| 1100 | CC stretching | Lipids and proteins |
| 1200 | C-C aromatic | Nucleic acids |
| 1250 | Amida III | Collagen I |
| 1300 | $CH_3$, $CH_2$ twisting | Collagen I |
| 1350 | $CH_2$ Bending | Collagen I |
| 1446 | $CH_2$ scissoring | Lipids e proteins |
| 1500 | C=C stretching | Proteins |
| 1580 | C-C stretching | Nucleic acids |
| 1660 | Amida I | Collagen I |
| 1730 | ? | Collagen III |

# FIGURES CAPTION

**Figure 1.** All Raman spectrum for the samples analysed.

**Figure 2.** A - Photomicrography - Box Plot MN. B - Photomicrography - Box Plot HFI.

**Figure 3.** Comparative analysis showing peaks of DNA and Collagen III.

**Figure 4.** Score plot obtained by using all wavelengths.

**Figure 5.** Score plot obtained by using a specific wavelength region.

**Figure 6.** Three-dimensional projection of the samples obtained by SIMCA method.

**Figure 7.** Plot of the distances among the classes of the samples studied.

# FIGURES

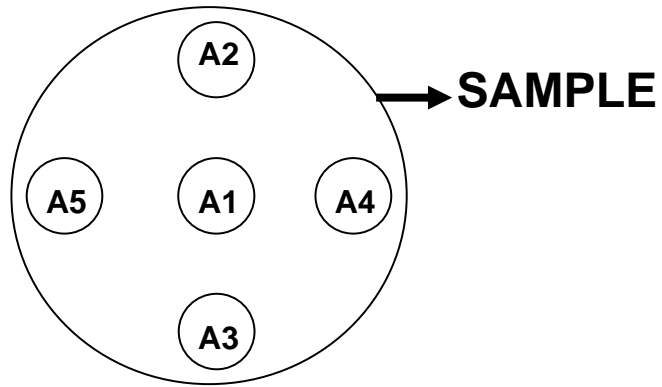

**Figure 1**

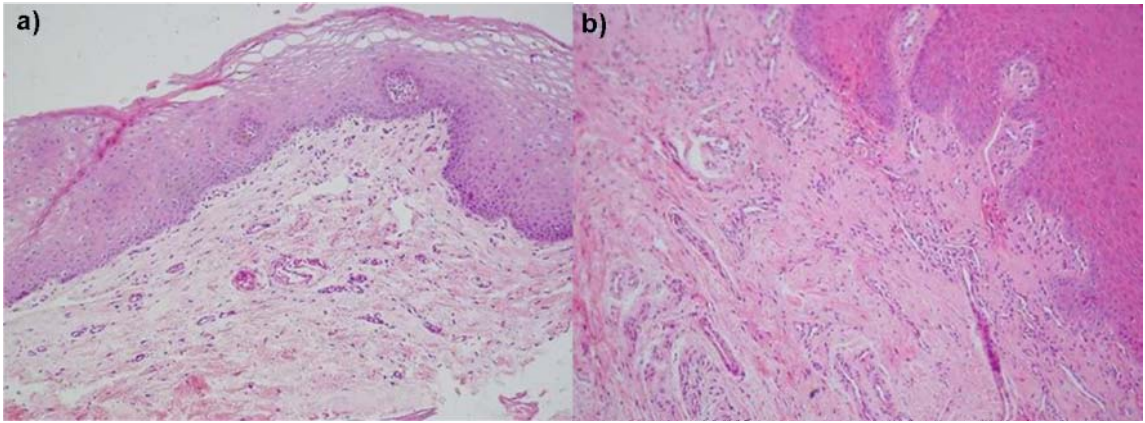

**Figure 2**

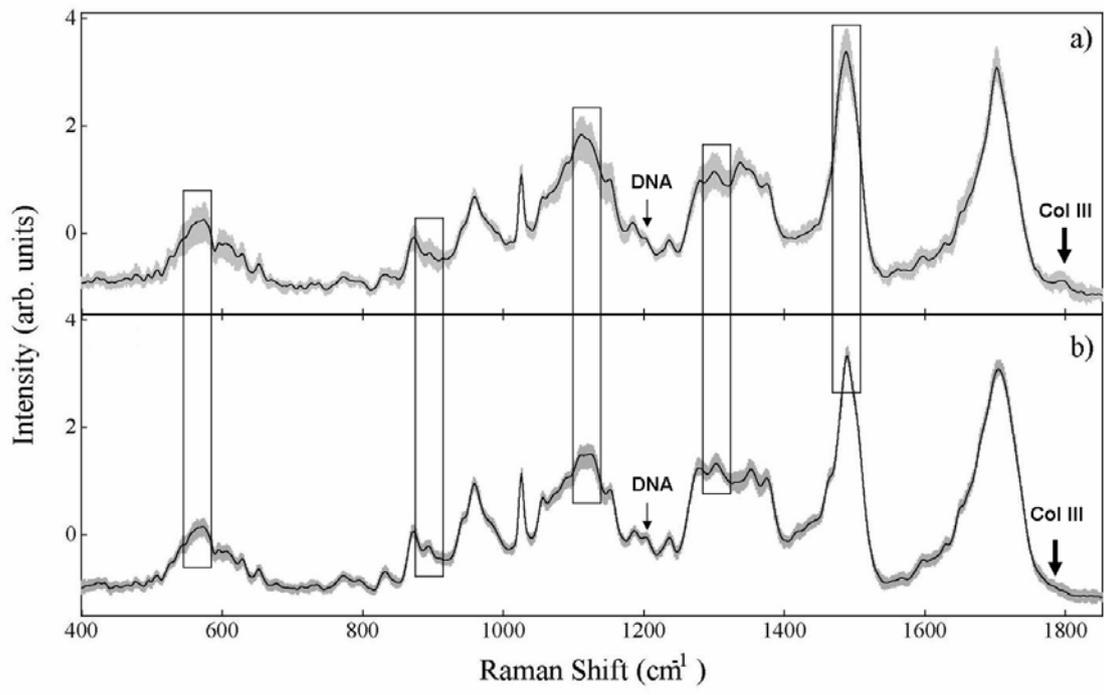

**Figure 3**

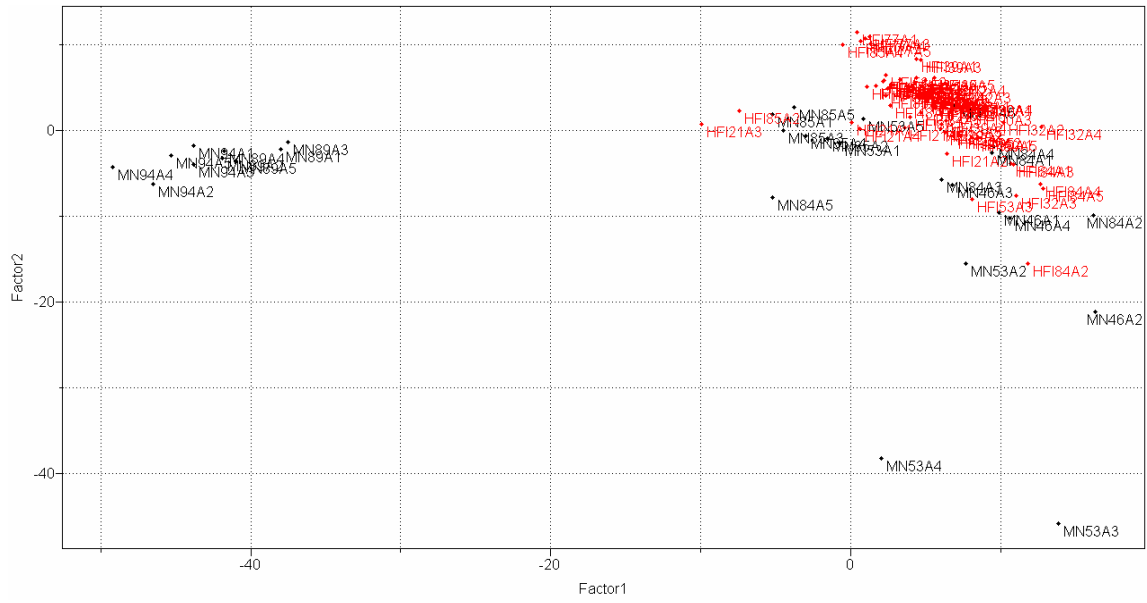

**Figure 4**

**Figure 5**

**Figure 6**

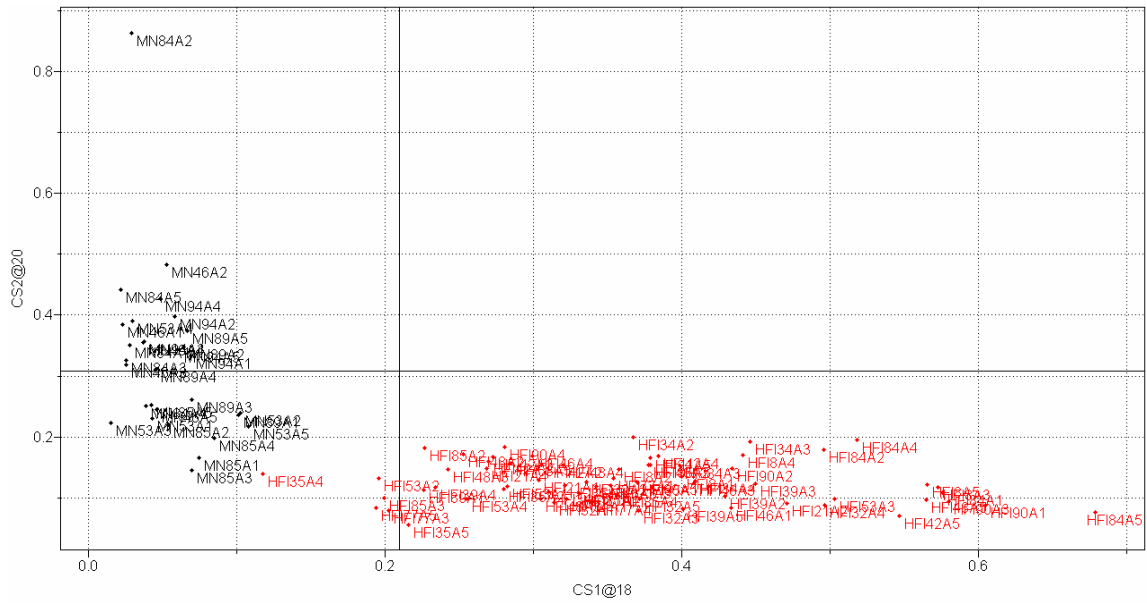

**Figure 7**